\documentclass[aoas,MSNbibl,nameyear,seceqn,rotating,dvips]{arximspdf}
\usepackage{graphicx}

\doi{10.1214/10-AOAS373}
\volume{5}
\issue{1}
\pubyear{2011}
\firstpage{201}
\lastpage{231}



\begin{document}
\begin{frontmatter}

\title{Bayesian methods to overcome the winner's curse~in genetic
studies\protect\thanksref{T1}}
\pdftitle{Bayesian methods to overcome the winner's curse~in genetic
studies}
\runtitle{Bayesian Methods for Winner's Curse}
\thankstext{T1}{Supported by a research grant from
the Canadian Institute of Health Research (CIHR).}

\begin{aug}
\author[A]{\fnms{Lizhen} \snm{Xu}\ead[label=e1]{lizhen@utstat.toronto.edu}},
\author[A]{\fnms{Radu V.} \snm{Craiu}\corref{}\thanksref{t1}\ead[label=e2]{craiu@utstat.toronto.edu}}
\and
\author[B]{\fnms{Lei} \snm{Sun}\thanksref{t1}\ead[label=e3]{sun@utstat.toronto.edu}%
\ead[label=u1,url]{http://www.foo.com}}
\thankstext{t1}{Supported by grants from the Natural
Sciences and Engineering Research Council of Canada (NSERC).}
\runauthor{L.  Xu,  R. V.  Craiu and L. Sun}

\affiliation{University of Toronto}

\address[A]{L. Xu\\
R. V. Craiu\\
Department of Statistics\\
University of Toronto\\
100 St. George Street\\
Toronto, Ontario M5S 3G3\\
Canada\\
\printead{e1}\\
\phantom{E-mail:\ }\printead*{e2}}

\address[B]{L. Sun\\
Dalla Lana School of Public Health \\
and Department of Statistics\\
University of Toronto\\
155 College Street\\
Toronto, Ontario M5T 3M7\\
Canada\\
\printead{e3}}
\end{aug}

\received{\smonth{9} \syear{2009}}
\revised{\smonth{5} \syear{2010}}

%
\begin{abstract}
Parameter estimates for associated genetic
variants, report ed in the initial discovery samples,
are often grossly inflated compared to the values
observed in the follow-up {replication} samples. This type of
bias is a consequence of the sequential procedure {in which the
estimated effect of an associated
genetic marker must first pass a stringent significance threshold.}
We propose a hierarchical Bayes method in which a spike-and-slab prior is
used to account for the possibility that the significant test result
may be due to chance.
We examine the robustness of the method using different priors
corresponding to
different degrees of confidence in the testing results and propose a
Bayesian model
averaging procedure to combine estimates produced by different models.
The Bayesian estimators yield
smaller variance compared to the conditional likelihood estimator and
outperform the latter in studies with low power.
We investigate the performance of the method with simulations and
applications to four real data examples.
\end{abstract}


\begin{keyword}
\kwd{Association study}
\kwd{Bayesian model averaging}
\kwd{hierarchical Bayes model}
\kwd{spike-and-slab prior}
\kwd{winner's curse}.
\end{keyword}

\end{frontmatter}

\section{Introduction}
Parameter estimates such as odds ratios (OR) for an associated genetic
variant (e.g., SNP, Single-Nucleotide Polymorphism), reported
{from the same discovery samples that were initially used to declare
statistical significance},
are often grossly inflated compared to the values
observed in the follow-up replication samples [e.g., \citet
{Nair2009ys}]. This type of
bias is a consequence of {using the same data for both model selection
and parameter estimation}, because a declared
associated variant must pass a stringent significance threshold.
This phenomenon is also known as the Beavis effect [\citet
{Xu2003ys}] or
the winner's curse [\citet{Zollner2007ys}] in the biostatistics
literature.

The winner's curse has recently gained much attention in genetic
studies, because 
{it has been} recognized as one of the major
contributing factors to the failures of many attempted replication
studies [e.g., \citet{Ioannidis2009ys}]. For example, five
\textit{Nature
Genetic} publications in the first three months of 2009 acknowledged
the effect of the winner's curse
[e.g., \citet{Nair2009ys}]. In their recent \textit{Nature Review}
paper, \citet{Ioannidis2009ys} dedicated a section to the
winner's curse and emphasized that ``\textit{the magnitude of the
winner's curse is inversely related to the power of the study. In
typical circumstances, for 10\% power, the inflation of an additive
effect could be approximately 60\%$\ldots.$ For small effects [anticipated
for susceptibility loci associated with complex diseases/traits],
even large meta-analyses could be grossly under-powered and emerging
associations could be considerably inflated. For rare variants, the
power can be~$<$1\%}.''

Some authors [e.g., \citet{Goring2001ys}] have argued that reliable
parameter estimates 
{can be obtained only} from an independent sample.
However, collecting additional samples could be undesirable due to,
for example, time and budget constraints as well as
concerns over population heterogeneity and sampling differences. Two
categories of methods were subsequently proposed to correct for the
selection bias using the original samples only: the model-free
resampling based methods [\citet{Sun2005ys}; \citet{wsb};
\citet{yuetal}; \citet{jeff2007ys}] and
the likelihood based methods
[\citet{Zollner2007ys}; \citet{ghosh2008ys}; \citet{Zhong2008ys}; \citet{xiao}]. Both
types of approaches were shown to substantially reduce the
estimation bias in relatively small samples, and comparable
performances were observed by
\citet{faye2009}. However, one caveat is that the variances of
the proposed estimators in both categories are considerably higher
than the original na\"{\i}ve estimator and lead to highly variable
estimates of the sample size needed for replication studies. {Although
the increased variability is expected, due to the bias-variance
trade-off, it may be too high to provide practical design recommendations}.
For example, Figure 4 of
\citet{Zollner2007ys} shows that the bias-adjusted sample size
estimates range from $\sim$500 to $\sim$100,000 compared to the
actual required sample size of 1,261 for a successful replication
study ($\alpha=10^{-6}$,  power${} = 80\%$).

Motivated by the above observations and the fact that some form of
prior information is often available in genetic studies, we propose
here a Bayesian framework to further reduce the bias and decrease
the variability in the estimates. In particular, we focus on the
OR estimates from genome-wide association studies (GWAS) via
logistic regression analyses of case-control disease status, because
most of the current genetic mapping studies adopt the case-control GWAS
design. We first describe the statistical model in Section~\ref{sec2}. We
prove in Section~\ref{sec3} that, conditional on statistical significance, there
are no unbiased estimators for the log OR. We
present the Bayesian methodology in Section~\ref{sec4} with detailed
discussions on the prior specifications and the advantages of model
averaging. We assess the performance of the proposed methods in
Section~\ref{sec5} via extensive simulation studies {under a general normal
model and specific genetic models. We demonstrate the utility of our
methods in
Section~\ref{sec6} with applications to four different association studies,
including a candidate gene study and three GWAS of either binary
case-control or quantitative outcomes.}
Our concluding remarks are in Section~\ref{sec7}.

\section{The statistical model}\label{sec2}

Let $\beta$ refer to the true log Odds Ratio (OR), the parameter of
interest, for the risk allele of an associated SNP, and $Z$ the
statistic of the corresponding association test. Following \citet
{ghosh2008ys},
we assume that $Z$ is asymptotically normally distributed and has
the form
\[
Z=\frac{\widehat{\beta}}{\widehat{SE} (\widehat{\beta})} \sim N
\biggl(\frac{\beta}{SE (\hat\beta)}, 1 \biggr),
\]
where $\widehat{\beta}$ is the estimate for $\beta$ from the
logistic regression, $\operatorname{logit} (E[Y]) = \alpha+ \beta X$, in
which the response variable $Y$ is the affection
status of a sample (0 $=$ unaffected and 1 $=$ affected by the
disease of interest) and the predictor $X \in\{0,1,2\}$ is the SNP
genotype coded
additively ($X$ represents the number of copies of the risk allele).
Other covariates may be also included in the model. Without loss of
generality, we assume
that the minor allele is the risk allele and the alternative of
interest is one-sided, that is, $H_0\dvtx \beta=0$ vs.  $H_1\dvtx \beta>
0$. The association test in this case is based on the Wald test,
and if the null hypothesis is rejected, the standard practice is to
directly use the $\widehat{\beta}$ from the logistic regression as the
estimate for $\beta$.

 The above estimation procedure is essentially the same as
the familiar practice of population mean estimation in the following
more general statistical setup.  Assuming that $n$ i.i.d. samples,
$\{X_1, \ldots, X_n\}$, were collected from a normal population with mean
$\mu$ and variance $\sigma^2$, a significance test is first conducted for
$H_0\dvtx \mu=0$ vs. $H_1\dvtx \mu>0$ based on the statistic,
$T_n={\frac{\overline{X}}{S/\sqrt{n}}}$, which follows
$N ({\frac{\mu}{\sigma/\sqrt{n}}}, 1 )$, where
$\overline{X}$ and $S$ are the sample mean and standard
deviation. The sample mean $\overline{X}$, calculated from the {\it
same} sample,
is subsequently used as an estimate for $\mu$, without adjusting for
the fact that
the null hypothesis was rejected (i.e., $T_n>c$, where $c$ is the
critical value corresponding to type I error rate $\alpha$) and that
estimation is performed for samples with positive
findings only.
Note that, in our simplified model, although $E[\overline{X}]=\mu$,
the conditional mean $E[\overline{X} | \overline{X} > (c S/\sqrt{n})
]$ is strictly greater than $\mu$, unless the
power of the test is 100\%. Thus, this na\"ive estimate, $\overline
{X}$, is upward biased. The amount of bias is inversely
proportional to the power as was first demonstrated by
\citet{Goring2001ys} in genome-wide linkage analyses and later by
\citet{Garner2007ys} for genome-wide association studies.
The likelihood based methods proposed by \citet{ghosh2008ys} and others
propose to correct for this selection bias by calculating
the maximum likelihood estimate (MLE) of $\mu$ from the correct
conditional likelihood. In this setting,
%
\begin{equation}\label{eqn:pdf}
P (\mathbf{X}|\mu, \sigma^2, T_n>c)=\prod_{i=1}^n
\frac{(1/\sqrt{2\pi\sigma^2})\exp[-{ (X_i-\mu
)^2/2\sigma^2}]}{1-\Phi(c-{\mu/(\sigma/\sqrt{n})})},
\end{equation}
where $\Phi$ is the cumulative distribution function (c.d.f.) of the
standard normal distribution.

{Although the above normal model is a conceptual one, it connects
directly with the logistic model used for case-control association
studies. Specifically, $\beta$ (the true log OR)
corresponds to $\mu$ (the normal population mean), $\widehat{\beta}$
(the na\"ive estimate)
corresponds to the statistic $\overline{X}$, and $\widehat{\mathit{SE}}
(\widehat{\beta})$ corresponds to $S/\sqrt{n}$. In the following
development of the bias correction Bayesian methods, we choose to focus
on the normal model for a number of reasons. The key factor that
influences the selection bias is the power of the association test,
which depends on the noncentrality parameter, $\beta/\mathit{SE} (\hat\beta
)$. In practice, $\beta$ is the true log OR, but $\mathit{SE} (\hat\beta)$ is
a complex function of multiple components including the prevalence of
the disease in the population, the disease model (e.g., additive,
dominant or others), the minor allele frequency of the SNP, the sample
size and the significance threshold used [\citet{Slager2001ys}].
The normal model allows us to concisely control the main factor of
interest, the power of the association test, in the simulation studies,
by fixing the normal population mean ($\mu\leftrightarrow\beta$, the
log OR) and considering practically meaningful ranges of significance
threshold value, power and sample size ($n$), which in turn determine
the normal population variance [$\sigma$, and $\sigma/\sqrt{n}
\leftrightarrow \mathit{SE} ( \hat\beta)$]. Moreover, this conceptual normal
model also covers association analyses of quantitative outcome, Y, for
which a linear regression model is typically used, for example, $E[Y] =
\alpha+ \beta X$. In that case, the population mean $\mu$ in the
conceptual normal model represents the regression coefficient
$\beta$. In Section~\ref{sec6} we show how our Bayesian methods built upon this
conceptual normal model can be applied to published association studies
for which only the OR (or the regression coefficient), the association
$p$-value, the sample size and the significance threshold were available.}

{In the following, we first show that there are no unbiased estimators
for the population mean conditionally on the significance of the
corresponding hypothesis test. We then proceed with the development of
a catalogue of Bayesian estimators and the evaluation of their
performance via simulation and application studies.}

\section{Lack of unbiased estimators for $\mu$} \label{sec3}{}

Ghosh, Zou and Wright (\citeyear{ghosh2008ys}) and other authors have demonstrated that the MLE
from the correct conditional likelihood could substantially reduce
the bias. However, they also observed via simulation studies that
the conditional MLE tends to over-correct for large $\mu$ and
under-correct for small $\mu$. {\citet{stallard2008ys} showed
that there is no conditional
unbiased estimators for the effect of treatment A from a sample that
was first used to select treatment A over B, that is, conditioning on
the fact that the sample effect of treatment A was larger than that
of treatment B. Although previous authors [\citet{Zhong2008ys};
 \citet{Bowden2009ys}] discussed that a similar argument can be used in the
case considered here,
below we provide a formal proof to show that there are no unbiased
conditional estimators for the population mean $\mu$ even when the
population variance $\sigma^2$ is known.}

Because $T_n$ is a sufficient statistic for $\mu$ when $\sigma$ is
known, the completeness of the normal family of
distributions implies that we can
restrict the search for unbiased estimators of
$\frac{\mu}{\sigma/\sqrt{n}}$ to functions of $T_n$.
Now suppose that some function $h (T_n)$ is an unbiased
estimator of $\frac{\mu}{\sigma/\sqrt{n}}$ conditional on the
statistical significance, that is, $T_n>c$.
Let $g (T_n)=\{T_n-h (T_n)\}$,
then
\begin{eqnarray*}
E[g (T_n)|T_n>c]&=&E[T_n|T_n>c]-E[h (T_n)|T_n>c]
\\
&=&\int_c^\infty\!T_n\frac{\phi(T_n-{\mu/(\sigma/\sqrt
{n})})}{1-\Phi(c-{\mu/(\sigma/\sqrt{n})})}\, d (T_n)
-\frac{\mu}{\sigma/\sqrt{n}}
\\
&=&\frac{1}{B}\int_{c-{\mu/(\sigma/\sqrt{n})}}^\infty
 \biggl(z+\frac{\mu}{\sigma/\sqrt{n}}\biggr)\phi(z)\, dz
-\frac{\mu}{\sigma/\sqrt{n}}
\\
&=&\frac{1}{B}\biggl[\int_{c-{\mu/(\sigma/\sqrt{n})}}^\infty z
\cdot e^{-{z^2/2}} \,dz
+B\cdot\frac{\mu}{\sigma/\sqrt{n}}\biggr]-\frac{\mu}{\sigma/\sqrt{n}}
\\
&=&\frac{1}{B}\biggl[\phi\biggl(c-\frac{\mu}{\sigma/\sqrt{n}}\biggr)+B\cdot\frac
{\mu}{\sigma/\sqrt{n}}\biggr]-\frac{\mu}{\sigma/\sqrt{n}}
\\
&=&\frac{\phi(c-{\mu/(\sigma/\sqrt{n})})}{1-\Phi(c-
{\mu/(\sigma/\sqrt{n})})},
\end{eqnarray*}
where $B=1-\Phi(c-\frac{\mu}{\sigma/\sqrt{n}})$.

Thus, we have
%
\begin{equation}
\int_c^\infty g (T_n)\frac{\phi(T_n-{\mu/(\sigma/\sqrt
{n})})}{1-\Phi(c-{\mu/(\sigma/\sqrt{n})})}\, dT_n
=\frac{\phi(c-{\mu/(\sigma/\sqrt{n})})}{1-\Phi(c-
{\mu/(\sigma/\sqrt{n})})},
\end{equation}
which implies
%
\begin{equation}\label{eqn:unbiased}
\int_c^\infty g (T_n)\phi\biggl(T_n-\frac{\mu}{\sigma/\sqrt{n}}\biggr) \,dT_n
=\phi\biggl(c-\frac{\mu}{\sigma/\sqrt{n}}\biggr).
\end{equation}

Now, let $\delta_c (y)$ be the Dirac delta function defined for
$y\geq c$ such that it is equal to 0 for all $y$ greater than $c$ and
$\int_c^\epsilon \delta_c (y)\, dy=1$ for all
$\epsilon>0$. It is easy to see that a solution to
equation (\ref{eqn:unbiased}) is $g (T_n)=\delta_c (T_n)$. By the
completeness of the normal distribution, the solution $g (T_n) \cdot
\mathbf{1}_{\{T_n>c\}}$ is
unique almost everywhere. Thus, $h (T_n) \cdot\mathbf{1}_{\{T_n>c\}
}=T_n\cdot\mathbf{1}_{\{T_n>c\}}$\vspace{2pt}
holds almost everywhere. Hence, $T_n$ is also an unbiased
estimator for $\frac{\mu}{\sigma/\sqrt{n}}$.
However, $T_n\cdot\mathbf{1}_{\{T_n>c\}}$ has an upward bias\vspace*{-4pt} equal to
$\frac{\phi(c-{\mu/(\sigma/\sqrt{n})})}{1-\Phi(c-{\mu/(\sigma/\sqrt{n})})}$.\vspace{5pt}
Therefore, we conclude that there are no unbiased estimators of
$\frac{\mu}{\sigma/\sqrt{n}}$ and hence no unbiased estimators of
$\mu$.

\section{Bayesian bias correction}\label{sec4}

\subsection{Prior specification}
The possible available prior information for genome-wide association
studies (GWAS) is diverse due to, for example, results from previous genome-wide
linkage analyses or candidate studies, or biological evidence
on the SNPs. One common theme, however, is the anticipated low power
of the GWAS and the well-acknowledged fact that an apparent
significantly associated SNP could be a false positive
[\citet{Ioannidis2009ys}]. Thus, the performance of the proposed
Bayesian methods is assessed in this context, although the practical
implementation of the methods could be study specific depending on
the type of the available prior.

The Bayesian paradigm allows us to incorporate in our model the prior
belief that the
significance of the effect observed \textit{may be due to chance}.
Mathematically, this belief can be
modeled using a spike-and-slab prior which is essentially a mixture
between a discrete probability with mass at
zero and a continuous density $f$ with support on the positive real line
\[
p (\mu|\xi)= \xi\delta_{\{0 \}} (\mu)+ (1-\xi) f (\mu)
\nonumber,
\]
where $\xi$ is either constant or a hyperparameter in the model.

The spike-and-slab priors have a long history in the Bayesian literature
on variable selection and shrinkage estimation, for example,
\citet{Box1986ys}, \citet{Mitchell1988ys},
\citet{George1993ys}, \citet{Chipman1996ys}, \citet
{Clyde1996ys},
\citet{Geweke1996ys}, and \citet{Kuo1998ys}. A recent theoretical
study by \citet{Ishwaran2005ys} discusses the similarities between
Bayesian procedures using the spike-and-slab priors and frequentist
procedures.

\begin{figure}

\includegraphics{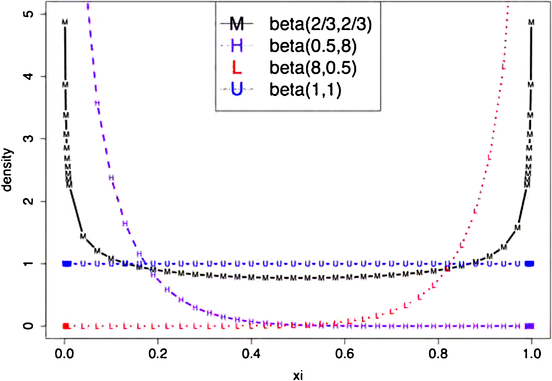}

\caption{Density of the prior $\operatorname{Beta} (a, b)$ for $\xi$
with different choices of $a$ and $b$.}
\label{fig1}
\end{figure}

We treat $\xi$ as a hyperparameter with a
Beta distribution, $\xi\sim\operatorname{Beta} (a, b).$
The parameters $a, b$ reflect our degree of prior belief in $\mu=0$
(false positive) versus $\mu> 0$ (true positive). If we set
$a=b=1$, then $p (\xi|a=1, b=1)$ is the $\operatorname{Uniform}(0,
1)$ density, which
implies that we do not favor, a priori, any region of $ (0, 1)$. This
could be considered the ``noninformative'' prior for $\xi$.
The choice $a=2/3$ and $b=2/3$ corresponds to our belief
in two extreme outcomes: \ $\xi$ is either close to 0 (believing in
true positive, $\mu>0$)
or close to 1 (believing in false positive, $\mu=0$).
Smaller values for $a$ and larger values for $b$, say, $a=0.5$ and $b=8$,
lead to a higher prior confidence that the signal is real.
Similarly, larger values for $a$ and smaller values for $b$, say,
$a=8$ and $b=0.5$, correspond to prior skepticism regarding the observed
association between the significant SNP and the trait of
interest.
Figure \ref{fig1} shows the Beta
distribution of $\xi$ for different values of $a$ and $b$.

Although we focus on
Beta(0.5, 8), and Beta(8, 0.5) in evaluating
the performance of the proposed Bayesian methods,
we conducted additional simulations to study the model's robustness to
the choice of priors.
Simulation results included in the supplementary material indicate
that other values for $a$ and $b$ [e.g., Beta(0.5, 16) or Beta(4,
0.5)]
that preserve the L-shaped or the ``inverse'' L-shaped density, as seen
in Figure~\ref{fig1},
produce very similar inferences.

In the existing likelihood approaches the sample variance, $S^2$, is
typically used to estimate $\sigma^2$ [\citet{ghosh2008ys}].
Although the variance estimator has relatively high precision in large
samples, it could be subject to the selection bias in small samples
[\citet{faye2009}].
Therefore, we adopt an empirical Bayes prior for $\sigma^2$ in which
the hyperparameters of the inverse gamma distribution, $\alpha_1$ and
$\alpha_2$, are chosen so that
the a priori mean of $\sigma^2$ is equal to $S^2$, the sample
variance, but the prior variance of $\sigma^2$ is equal to $200$. We
note that additional simulations with more certainty about $\sigma^2$
(prior variance of $\sigma^2$ as small as 10) or less certainty (as
large as 1000) produce very similar results.

We use $\operatorname{Uniform}(0, A)$ to specify $f (\mu)$, the
density function for
the continuous component of the prior for $\mu$, the log OR,
where $A$ represents the upper bound of log OR.
However, in this parametrization the estimator is very sensitive to the
choice of
$A$. To show this, let $Z$ be the latent mixture indicator so that
$Z=0$ if the significant SNP is a
false positive ($\mu=0$) and $Z=1$ for a true positive ($\mu>0$). It
is not difficult to see that
\[
Z|\vec{X}, \xi, \mu, \sigma^2= \cases{
0, &\quad\mbox{with probability } $\dfrac{p_0}{p_0+p_1}$,\cr
1, &\quad\mbox{with probability } $\dfrac{p_1}{p_0+p_1}$,
}
\]
where $\vec X= \{X_1, \ldots, X_n\}$ and
\begin{eqnarray*}
p_0 &=&
\frac{\xi}{1-\Phi(c)},
\\
p_1&=&\frac{1 }{ A} \times\frac{ (1-\xi) \exp\{-{(1/(2\sigma^2))} (n\mu
^2-2\mu\sum_{i=1}^n X_i)\}
}{1-\Phi(c-{\mu/(\sigma/\sqrt{n})})}.
\end{eqnarray*}
Thus, depending on the value of $A$, $p_1$ can be made arbitrarily
small regardless of the data available. This can influence dramatically
(even for $A=2$) the performance of the computational algorithm used to
obtain the posterior distribution of interest (described in Section
\ref{MCMC}).
One simple method to circumvent this problem is to use the
reparametrization $\theta=\mu/A$ which dissolves the influence of $A$
on~$p_1$.
Therefore, the proposed Bayesian method has the
following hierarchical prior structure:
%
\begin{eqnarray} p (\theta|\xi)&=&\xi g_0 (\theta)+ (1-\xi)g_1
(\theta),
\\
\xi&\sim&\operatorname{Beta} (a, b),\nonumber
\\
\sigma^2 &\sim&\operatorname{Inv\mbox{-}Gamma} (\alpha_1, \alpha_2),\nonumber
\end{eqnarray}
where
$\alpha_1=S^4/200 +2$, and
$\alpha_2=S^6/200 +S^2$,
$S$ is the sample standard devia\-tion,~$g_0 (\theta)=\delta_{\{0\}}
(\theta)$ and $g_1 (\theta)$ is the density of Uniform(0, 1).

{In the actual implementation, we use $A=2$ to reflect
the known maximum log OR of SNPs identified for complex diseases and
traits. For example,
the truly associated SNP in the well-known major histocompatibility
complex (MHC) region has perhaps the highest genetic effect observed
to date, with a log OR of $\log (5.49)=1.7$ [WTCCC (2007)].
We note that additional simulations showed that, as long as
the reparametrization $\theta=\mu/A$ is used,
results remain largely the same for higher upper bounds (e.g., $A=6$
corresponding to a maximum OR${}\approx400$). Applications in Section 6
also demonstrate the
robustness of the model when it was applied not only to case-control
data but also to an association study of a quantitative outcome.}

\subsection{Posterior distribution}

The joint prior distribution for
$ (\theta, \xi)$ is
\begin{eqnarray}
\label{pr1}
p (\theta, \xi)&=& p (\theta|\xi)p (\xi)\nonumber
\\[-8pt]\\[-8pt]
&=&
\xi g_0 (\theta) \xi^{a-1} (1-\xi)^{b-1}+
(1-\xi) g_1 (\theta) \xi^{a-1} (1-\xi)^{b-1}.\nonumber
\end{eqnarray}

Conditional on $Z$, the sampling distribution is
\begin{eqnarray*}
&&P (\vec X|\theta, \sigma^2, Z, T_n>c)
\\
&&{}\qquad \propto
(1/\sigma)^n \biggl(\frac{\exp\{-\sum_{i=1}^n {X_i^2/(2\sigma^2)}\}
}{1-\Phi(c)}\biggr)^{1-Z}
\\
&&{}\qquad\quad \times\biggl(\frac{\exp\{-\sum_{i=1}^n { (X_i-2\theta)^2/(2\sigma^2)}\}
}{1-\Phi(c-{2\theta/(\sigma/\sqrt{n})})}\biggr)^Z.
\end{eqnarray*}
If $Z$ were observed, the posterior distribution for the
vector $ (\theta, \xi, \sigma^2)$ would be
%
\begin{eqnarray}
\label{eq:post}
&& p (\theta, \xi, \sigma^2|\vec X, Z, T_n>c)\nonumber
\\
&&\quad{} \propto
p (\vec X, Z|\theta, \sigma^2, T_n>c)p (\theta|\xi)p (\xi)p
(\sigma^2) \nonumber
\\
&&\quad{} \propto (1/\sigma)^n \biggl(\frac{\exp\{-\sum_{i=1}^n
{X_i^2/(2\sigma^2)}\}\xi}{1-\Phi(c)}\biggr)^{1-Z}
\\
&&{}\qquad\times\biggl(\frac{\exp\{-\sum_{i=1}^n { (X_i-2\theta)^2/(2\sigma^2)}\}
(1-\xi)}
{1-\Phi(c-{2\theta/(\sigma/\sqrt{n})})}\biggr)^Z \nonumber
\\
&&{}\qquad \times
\xi^{a-1} (1-\xi)^{b-1} \biggl(\frac{1}{\sigma^2}\biggr)^{\alpha_1+1}\exp\{
-{\alpha_2/\sigma^2} \}\nonumber
\end{eqnarray}
for $\theta, \xi\in[0, 1]$, $\sigma>0$
(detailed derivation provided in the \hyperref[sup]{Supplementary }
\hyperref[sup]{material}).
We note that the posterior distribution specified in equation (\ref
{eq:post}) depends on the data only through the sufficient statistics
for $ (\mu, \sigma^2)$, $D_n= (\sum X_i, \sum X_i^2)$.
This is particularly useful in practice when the original
sample-specific data $\vec X$ are not available, but the sufficient
statistics are provided or could be inferred from typically reported
quantities such as the sample size, the observed OR and association
$p$-value, and the significance threshold used.

\subsection{Sampling from the posterior distribution}\label{MCMC}

The latent variable $Z$ is unobservable in practice, so equation (\ref
{eq:post}) cannot be used directly to study the characteristics of the
posterior distribution, $\pi(\theta, \xi, \sigma^2)=p (\theta,
\xi, \sigma^2|D_n,\break T_n>c)$.
The traditional approach in this type of situation is to use Markov
chain Monte Carlo (MCMC) techniques to sample from $\pi$. The
posterior distribution has a mixture form for which the Data
Augmentation algorithm of \citet{tann-wong} has been proven
extremely efficient [see also \citet{van}].
The algorithm relies on sampling alternatively from the distribution of
$Z|D_n, \theta, \xi, \sigma^2$ and
$\theta, \xi, \sigma^2|Z, D_n$. More precisely, at iteration $t$ we
carry out the following steps:

\textit{Step} 1. Sample $Z_t \in\{0, 1\}$ given
$\xi_{t-1}$, $\theta_{t-1}$ and $\sigma^2_{t-1}$ from the conditional
distribution
\[
Z_t|\xi_{t-1}, \theta_{t-1}, \sigma^2_{t-1}= \cases{
0, &\quad\mbox{with probability }$\dfrac{p_0}{p_0+p_1}$,\cr
1, &\quad\mbox{with probability }$\dfrac{p_1}{p_0+p_1}$,
}
\]
where
\begin{eqnarray*}
p_0 &=&
\frac{\xi_{t-1} }{1-\Phi(c)},
\\
p_1&=&\frac{ (1-\xi_{t-1}) \exp\{-{(1/(2\sigma^2_{t-1}))} (4n\theta
_{t-1}^2-4\theta_{t-1}\sum_{i=1}^n X_i)\}
}{1-\Phi(c-{2\theta_{t-1}/(\sigma_{t-1}/\sqrt{n})})}.
\end{eqnarray*}

\textit{Step} 2.
(i) If $Z_t=0$, sample
\begin{eqnarray*}
\xi_t &\sim&\operatorname{Beta} (a+1, b),
\\
\sigma^2_t &\sim& p(\sigma^2|D_n) \propto\biggl(\frac{1}{\sigma
^2}\biggr)^{n/2+\alpha_1+1}\exp\biggl\{-\frac{1}{\sigma^2} \biggl(\alpha_2+\frac
{\sum_{i=1}^n X_i^2}{2} \biggr) \biggr\},
\end{eqnarray*}
which is the inverse gamma distribution with shape parameter equal to
$\frac{n}{2}+\alpha_1$,\vspace*{-2pt} and scale parameter equal to
$\alpha_2+\frac{\sum_{i=1}^n X_i^2}{2}$.
We also set $\mu_t=\theta_t=0$.

(ii) If $Z_t=1$, sample
\begin{eqnarray*}
\xi_t &\sim&\operatorname{Beta} (a, b+1),
\\
\theta_t &\sim& p (\theta|D_n, \sigma_{t-1}) \propto\frac{\exp\{-
{2n\theta^2/\sigma_{t-1}^2}-{2\theta\sum_{i=1}^n
X_i/\sigma_{t-1}^2}\}
}{
1-\Phi(c-{2\theta/(\sigma_{t-1}/\sqrt{n})})}\mathbf{1}_{ (0,
1)} (\theta),
\\
\sigma^2_t&\sim& p (\sigma^2| (D_n, \xi_{t}, \theta_t) \propto
\frac{\exp\{-{1/(2\sigma^2)} (\sum_{i=1}^n X_i^2+4n\theta
_t^2-4\theta_t\sum_{i=1}^nX_i)\}
}{
(1-\Phi(c-{2\theta_t/\sqrt{\sigma^2/n}}))}
\\
&&{}\hspace*{87pt}\times (\sigma^2)^{n/2+\alpha_1+1}\exp\{-\alpha_2/\sigma^2\}.
\end{eqnarray*}
The sampling of $\theta_t$ and $\sigma^2_t$ at step 2(ii) cannot be
carried out directly, so we apply a Metropolis--Hasting algorithm
[\citet{metrop}]. We use 20,000 iterations to obtain 15,000 posterior
samples, discarding the first 5000 ``burn-in'' samples.
The sample mean of the
above 15,000 posterior samples, $\overline\theta$, is used to
estimate the posterior mean $E[\mu|D_n, T_n>c]$. That is,
$\widehat{\mu}_B = 2 \overline\theta$, where the factor 2 is due to
the initial
reparametrization $\theta=\mu/A$ and $A=2$.
 (Additional simulations presented in the \hyperref[sup]{Supplementary material}
 show that running the chain longer or discarding
more ``burn-in'' samples provide similar results.)

\subsection{Bayesian Model Averaging (BMA)}

The Bayesian model averaging (BMA) is a coherent and conceptually
simple method devised to take into account the model uncertainty
[see \citet{BMA1999ys} and references therein]. For the problem
discussed here, the uncertainty is related to our lack of
information regarding the power of the test performed in the first
stage. If we knew, say, that the power of the test is high, then we
would be more confident that the signal detected is a true signal
and this would be reflected in our choice of the prior.
In the absence of such information, one could adopt the
BMA methodology to increase the robustness of the Bayesian
estimator.

In the BMA paradigm, assume that $\Delta$ is the quantity of
inferential interest for which a number of candidate models,
say, $M_1, \ldots, M_K$, are available. Given the prior probability for
each candidate model, $p (M_i), 1
\le i \le K$, the traditional
BMA method assigns the posterior distribution given data $D$ for~$\Delta$
%
\begin{equation}\label{Delta}
p (\Delta|D)=\sum_{k=1}^K   p (\Delta|M_k, D)p (M_k|D),
\end{equation}
where
\[
p (M_k|D)=\frac{p (D|M_k)p (M_k)}{\sum_{l=1}^K
p (D|M_l)p (M_l)}
\]
and
\[
p (D|M_k)=\int p (D|\theta_k, M_k)p (\theta_k|M_k)\, d\theta_k.
\]

In our setting, $K=2$ because only two models are considered.
Let $M_1$ be the model with prior $p (\xi)=\operatorname{Beta} (8, 0.5)$ (a
priori favors the belief that the initial discovery is a false positive)
and $M_2$ for $p (\xi)=\operatorname{Beta} (0.5, 8)$ (a priori favors the
belief that the initial discovery is a true positive).
To specify the values for $p (M_1)$ and $p (M_2)$, we utilize the
threshold value $c$ in the following fashion, $p (M_1)=e^{ (-c/2)}$
and $p (M_2)=1-e^{ (-c/2)}$. Thus, our prior belief in model $M_1$
(with higher density for false positive) decreases as the testing
threshold value increases at an exponential rate. The posterior
probabilities for the two models can be derived as
\[
p (M_i|D_n)=\frac{p (D_n|M_i)p (M_i)}{p (D_n|M_1)p (M_1)+p (D_n|M_2)p (M_2)},
\qquad i=1, 2.
\]
Thus,
%
\begin{equation}\label{model_posterior_ratio}
\frac{p (M_1|D_n)}{p (M_2|D_n)}=\frac{p (D_n|M_1)}{p (D_n|M_2)}
\cdot\frac{e^{ (-c/2)}}{ (1-e^{ (-c/2)})}.
\end{equation}

The direct computation, however, is difficult because the integral
\[
p (D_n|M)=\int\hspace{-2pt}\int_{ (\mu, \xi, \sigma^2)}
p (D_n|\mu, \xi, \sigma^2, M)p (\mu|\xi, M)p (\xi|M)p (\sigma
^2|M) \,d\mu \,d\xi
\]
cannot be calculated in a closed form. Note that
%
\begin{equation}\label{normal_const_id}
\qquad p (\mu, \xi, \sigma^2|D_n, M)=\frac{p (D_n|M, \mu, \xi, \sigma
^2)p (\mu|\xi, M)p (\xi|M)p (\sigma^2|M)}{p (D_n|M)},
\end{equation}
thus $p (D_n|M)$ can be viewed as the normalizing
constant of the posterior distribution $p (\mu, \xi, \sigma^2|D_n, M)$.
Therefore, the first ratio in (\ref{model_posterior_ratio}) is a ratio
of two normalizing constants for
two densities from which we can sample. The problem of estimating
ratios of two normalizing constants
has been discussed by, among others, \citet{meng1996ys} and
\citet{MR1647507}. We use the
bridge sampling method proposed by \citet{meng1996ys} to compute the
ratio in~(\ref{model_posterior_ratio}).

To compute (\ref{model_posterior_ratio}), let $r=p (D_n|M_1)/p (D_n|M_2)$,
$\omega= (\mu, \xi, \sigma^2)$, $\pi_i=p (\mu, \xi,\break \sigma
^2|D_n, M_i) $
and
$ q_i (\mu, \xi, \sigma^2)=p (D_n|M_i, \mu, \xi, \sigma^2)p
(\mu|\xi, M_i) p (\xi|M_i)p (\sigma^2|M_i), $ for $1\le i \le
2.$ Given $m=10{,}000$ samples
$\{ (\mu_{i1}, \xi_{i1}, \sigma^2_{i1}),\ldots, (\mu_{in_i},
\xi_{in_i}, \sigma^2_{i1})\}$
from each density $\pi_i$, we can approximate $r$ using the iterative
procedure of \citet{meng1996ys}. Specifically, after starting
with an initial estimate $\hat{r}^{ (0)}$, at the $ (t+1)$st iteration, we
compute
\begin{eqnarray}\label{iter_alpha}
\hat{r}^{ (t+1)}&=&\frac{{(1/m)}
\sum_{j=1}^{m}[{q_1 (\omega_{2j})/(s_1q_1 (\omega
_{2j})+s_2\hat{r}^{ (t)}q_2 (\omega_{2j}))}]}
{{(1/m)}
\sum_{j=1}^{m}[{q_2 (\omega_{1j})/(s_1q_1 (\omega
_{1j})+s_2\hat{r}^{ (t)}q_2 (\omega_{1j}))}]}\nonumber
\\[-8pt]\\[-8pt]
&\equiv&\frac{{(1/m)}\sum_{j=1}^{n_2}[
{l_{2j}/(s_1l_{2j}+s_2\hat{r}^{ (t)})}]}
{{(1/m)}\sum_{j=1}^{m}[{1/(s_1l_{1j}+s_2\hat{r}^{ (t)})}]},\nonumber
\end{eqnarray}
where $s_i=0.5$, and
$l_{ij}=\frac{q_1 (\omega_{ij})}{q_2 (\omega_{ij})}, $
for $1\le j \le m$, $1\le i \le2$. Note that $l_{ij}$ needs to be
computed only once at the
beginning of
the algorithm. The convergent value of $\hat{r}^{ (t)}$ is the one we
choose to estimate $r$.

In the current setting $l_{ij}$ is easy to compute since
\begin{eqnarray*}\label{l_ij}
l_{ij}&=&\frac{p (D_n|M_1, \mu_{ij}, \xi_{ij}, \sigma^2_{ij})p (\mu
_{ij}|\xi_{ij}, M_1)p (\xi_{ij}|M_1)p (\sigma^2_{ij}|M_1)}
{p (D_n|M_2, \mu_{ij}, \xi_{ij}, \sigma^2_{ij})p (\mu_{ij}|\xi
_{ij}, M_2)p (\xi_{ij}|M_2)p (\sigma^2_{ij}|M_2)}
\\
&=&\frac{p (\xi_{ij}|M_1)}{p (\xi_{ij}|M_2)}=\xi_{ij}^{7.5} (1-\xi
_{ij})^{-7.5}.
\end{eqnarray*}
From equations (\ref{Delta})
and (\ref{model_posterior_ratio}), we obtain the BMA estimator of
$\mu$,
%
\begin{equation}\label{eq:mu_BMA}
\hat\mu_{\mathit{BMA}}=\frac{\hat{r} e^{ (-c/2)}}{\hat{r} e^{ (-c/2)}+1-e^{
(-c/2)}}\hat\mu_1
+\frac{1-e^{ (-c/2)}}{\hat{r}e^{ (-c/2)}+1-e^{ (-c/2)}}\hat\mu_2,
\end{equation}
where $\hat\mu_1$ and $\hat\mu_2$ are the posterior means of $\mu$
obtained under models $M_1$ and $M_2$, respectively.

\section{Simulation study}\label{sec5}

We carried out two sets of simulations to examine the performances of
the Bayesian methods
and compared the results with those from the likelihood-based
estimators of \citet{ghosh2008ys}. The first set of simulations
used data generated from the normal model that was used to outline and
develop the Bayesian methods, and the second set used data simulated
from a case-control genetic model. The nine estimators examined are as follows:

\begin{description}
\item[N:] The na\"ive estimator ($\overline X$, the {\it
unconditional} MLE).
\item[MLE:] The \textit{conditional} MLE estimator based on equation
(\ref{eqn:pdf}), that is the $\beta_1$ estimator in
\citet{ghosh2008ys}.
\item[NMLE:] The mean of the Normalized Conditional Likelihood
estimator, that is, the $\beta_2$ estimator of
\citet{ghosh2008ys}.
\item[Ghosh:] The average estimator of MLE and NMLE, that is, the
$\beta_3$ estimator recommended by \citet{ghosh2008ys}.
\item[B.L:] The Bayesian estimator based on equation (\ref{eq:post})
when the prior for $\xi$ is $\operatorname{Beta} (8, 0.5)$ (the prior belief
is low power of the initial discovery study).
\item[B.H:] The Bayesian estimator based on equation (\ref{eq:post})
when the prior for $\xi$ is $\operatorname{Beta} (0.5, 8)$ (the prior belief
is high power of the initial discovery study).
\item[B.BMA:] The BMA estimator obtained by averaging the B.L and B.H
models, based on equation (\ref{eq:mu_BMA}).
\item[B.M:] The Bayesian estimator based on equation (\ref{eq:post})
when the prior for $\xi$ is $\operatorname{Beta} (2/3, 2/3)$ (the prior
belief is either low or high power).
\item[B.Unif:] The Bayesian estimator based on equation (\ref
{eq:post}) when the prior for $\xi$ is $\operatorname{Uniform}(0, 1)
$ (the
``noninformative'' prior).
\end{description}
Whenever an obtained estimate was negative, it was truncated to be zero
following the standard practice of
interpreting the ``flip--flop'' phenomenon occurring at the same SNP
in the same population [\citet{flip-flop}]. That is, a SNP is
found to be associated with the disease of interest in two independent
studies, but the risk allele is reversed (i.e., the allele that
increases the risk in one study is the protective allele that decreases
the risk in another study).

\subsection{Simulation set 1---normal model}
We considered a factorial design in which
the factors are the power of the association test, the type 1 error
rate and the sample size.
The power levels are
$\{5\%, 10\%, 20\%, $ $ 50\%, 99\%\}$, of which $99\%$ allows us to
investigate the asymptotic behavior
of the methods while 20\% or lower reflect the low power
anticipated for genome-wide association studies (GWAS). The type 1
error rates, $\alpha$,
are $\{0.05, 10^{-4}, 10^{-6}\}$, of which 0.05 is the typical choice
for a single SNP study, while the other two are suitable for
high-throughput GWAS depending on the density of the SNPs being
genotyped. The corresponding threshold values for the test statistics,
$c$, are $\{1.645, 3.719, 4.753\}$.
The true population mean is fixed at $\mu=0.095 = \log (1.1)$, and the
sample size ranges from $n=100$ to over 10,000 depending on the
combination of $\alpha$ and power.
The values of the these parameters then uniquely determine the
corresponding population variance,~$\sigma^2$. The details of each
simulation scenario are shown in Table~\ref{tab1}.

Under each simulation scenario, we began by generating 200 significant data
sets, that is, $X_i \sim N (\mu, \sigma^2), i=1, \ldots, n$, such that
the value of
the test statistic, $T_n={\frac{\overline{X}}{S/\sqrt{n}}}$, is
greater than $c$.
We then computed the nine estimates, \textbf{N}, \textbf{MLE}, \textbf{NMLE},
\textbf{Ghosh}, \textbf{B.L}, \textbf{B.H},
\textbf{B.BMA}, \textbf{B.M} and \textbf{B.Unif}, for each significant data set.

\begin{sidewaystable}
\tablewidth=\textwidth
\caption{Simulation scenarios for the normal model}\label{tab1}
\tabcolsep=3pt
\begin{tabular*}{\tablewidth}{@{\extracolsep{\fill}}lccccccccccccccc@{}}
\hline
&\multicolumn{3}{c}{\textbf{5\%}}&\multicolumn{3}{c}{\textbf{10\%}}&\multicolumn{3}{c}{\textbf{20\%}}&
    \multicolumn{3}{c}{\textbf{50\%}}&\multicolumn{3}{c@{}}{\textbf{99\%}} \\[-6pt]
    &\multicolumn{3}{c}{\hrulefill}&\multicolumn{3}{c}{\hrulefill}&\multicolumn{3}{c}{\hrulefill}&
    \multicolumn{3}{c}{\hrulefill}&\multicolumn{3}{c@{}}{\hrulefill} \\
 $\bolds{\alpha\backslash\mbox{power}}$   &  $\bolds{n}$& $\bolds{\sigma}$ & $\bolds{\sigma/\sqrt{n}}$
 &$\bolds{n}$ & $\bolds{\sigma}$ & $\bolds{\sigma/\sqrt{n}}$ &$\bolds{n}$& $\bolds{\sigma}$ & $\bolds{\sigma/\sqrt{n}}$
     &$\bolds{n}$& $\bolds{\sigma}$ & $\bolds{\sigma/\sqrt{n}}$ &$\bolds{n}$& $\bolds{\sigma}$& $\bolds{\sigma/\sqrt{n}}$ \\
     \hline
     \phantom{0}0.05&--&--&--&\phantom{0}100&2.623& 0.262& \phantom{0}200&1.678&0.119&$1000$&1.832&0.058&\phantom{0,}$5000$&1.697&0.024\\
     $10^{-4}$&1000&1.453&0.046& 2000&1.749&0.039& 3000&1.814&0.033& 5000&1.812&0.026& 10,000&1.577&0.016\\
     $10^{-6}$&2000&1.371&0.031&4000&1.736&0.027& 5000&1.723&0.024& 8000&1.793&0.020& 16,000&1.702&0.013\\
     \hline
\end{tabular*}
\tabnotetext[]{}{\textit{Notes}: Sample size   ($n$) and population standard error
($\sigma$) needed to obtain the desired power at the prespecified type 1 error rate   ($\alpha$) when population mean $\mu=0.0953=\log  (1.1)$.}
\end{sidewaystable}

\begin{figure}

\includegraphics{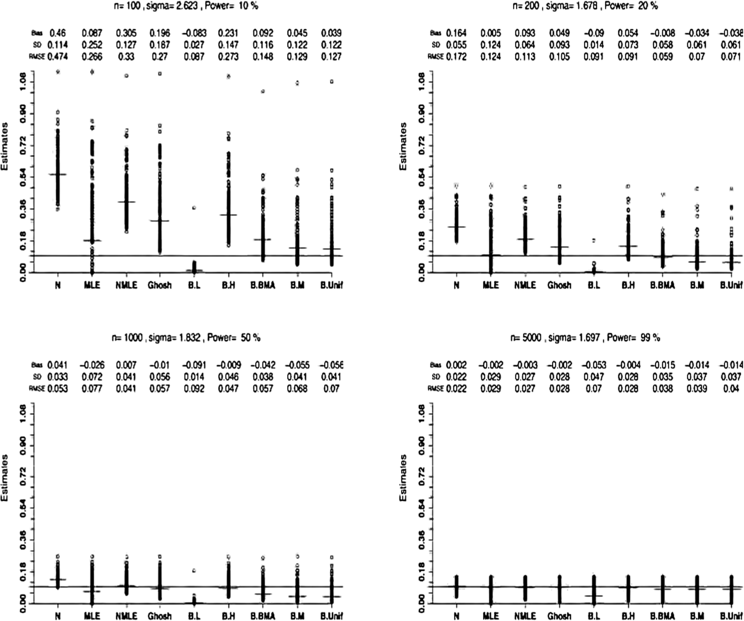}

\caption{Performance of the nine estimators under the normal
model with a type 1 error rate of 0.05. The population mean $\mu=\log
(1.1)=0.0953$ and power ranging from 10\%, 20\%, 50\% to 99\%. Details
of the simulating parameters are given in row 1 of Table~\protect\ref{tab1}. Each circle
represents an estimate, the horizontal is the averaged estimate over
200 simulated data sets, and the long horizontal line represents the
true value of $\mu$. The Bias, sample Standard Deviation (SD) and Root
Mean Squared Error (RMSE) are also provided for each estimator.}
\label{fig2}
\end{figure}

Figure~\ref{fig2} provides detailed results when the type 1 error rate is 0.05
and the simulating parameter values are those in row 1 of Table~\ref{tab1}.
These plots confirm that, in the
case of low power of the initial association study (e.g., 10\%), the
na\"ive estimator has a large upward bias. Even
in the moderately powered studies (e.g., 20\%), the na\"ive estimator
could considerably
overestimate the true effect size. Note that the two priors with opposite
degrees of belief in the significance of the effect, \textbf{B.L} and \textbf{B.H},
produce quite different results. The \textbf{B.L} estimator conservatively
shrinks the effect and, therefore,
it is more reliable in those cases when the effect is small or zero.
(See additional figures in \hyperref[sup]{Supplement}  for the case of no genetic
effect, i.e., the apparent association is a false positive.)
When the power of the test is relatively high (e.g., 50\%), \textbf{B.H}
outperforms the other estimators considered.
While it is clear that \textbf{B.L} and \textbf{B.H} are complementing each
other, \textbf{B.BMA}, designed to balance between \textbf{B.L} and {\bf
B.H}, performs
well in a variety of settings. The performances of the other two
estimators, \textbf{B.M} and \textbf{B.Unif}, are similar to one another but
inferior to \textbf{B.BMA}.
The natural implication is that putting equal prior weight on
$ (0, 1)$ is equivalent to putting equal weight on $\xi$ close to zero
or close to 1. As expected, when the power is very high (e.g., 99\%)
there is little bias in the na\"ive estimate; the other estimates also
converge to the true value with \textbf{B.L} lagging behind. This is due
to the strong skepticism embedded in the \textbf{B.L} model about the finding.

In most of the cases, the Bayesian estimators achieve the
anticipated reduction in bias as well as variance compared to the
likelihood based estimators,
\textbf{MLE}, \textbf{NMLE} and \textbf{Ghosh}. Of the three, we observed that
\textbf{Ghosh}
(i.e., the average of \textbf{MLE} and \textbf{NMLE}) performs the best,
confirming the conclusion of \citet{ghosh2008ys}. Therefore, in
what follows we focus on the comparison between \textbf{B.BMA} and \textbf{Ghosh}.

The advantage of \textbf{B.BMA} over \textbf{Ghosh} is especially obvious in
the low power studies. For example, when the power of the test is $10\%$,
the bias of \textbf{Ghosh} is 0.196, almost twice as big as 0.092 for \textbf{B.BMA}.
The sample standard deviation of the \textbf{Ghosh} estimate is 0.186
compared to 0.116 for the \textbf{B.BMA} estimate. The Root Mean Squared
Error (RMSE) for \textbf{B.BMA} is almost half that for \textbf{Ghosh} (0.148
vs. 0.273).
To formally assess the significance of the difference between {\bf
Ghosh} and \textbf{B.BMA},
we performed a matched-pair $t$-test based on 50 simulation runs, and we
obtained a $t$-statistic of $-$117.47 showing that the difference is significant.
As expected, the advantage dissipates and the two perform similarly
when the power of the initial association study increases.

As discussed by \citet{ghosh2008ys} and detailed in Section~\ref{sec2},
the main factor that influences the estimation bias is the power of the
association test which depends on the noncentrality parameter, $\mu/
(\sigma/\sqrt{n})$. Thus, although $\mu$ has the interpretation of
$\beta= \log \mathrm{OR}$ and was fixed at $\log (1.1)$, the results are
qualitatively similar for larger OR with smaller sample size or smaller
OR with larger sample size, as long as the ratio, $\mu/ (\sigma/\sqrt
{n})$, and the significance threshold value, $\alpha$, stay the same.

\begin{figure}

\includegraphics{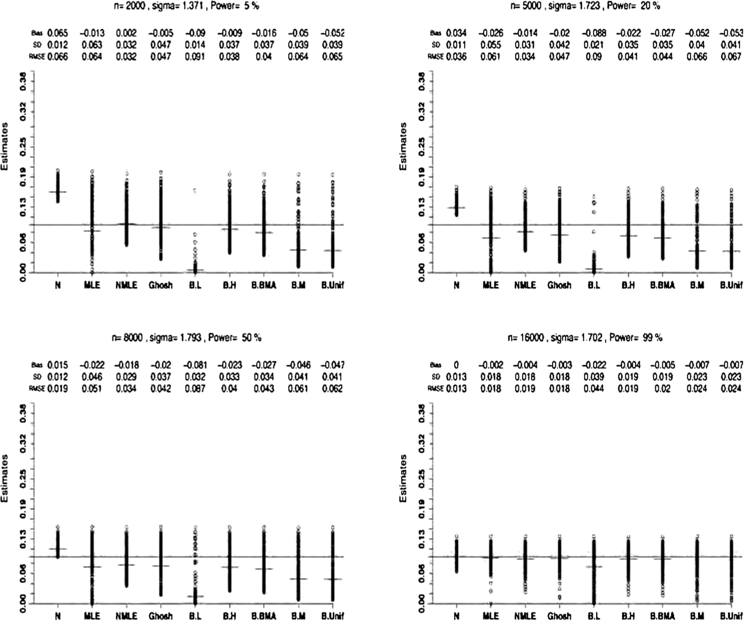}

\caption{Performance of the nine estimators under the normal
model with a type 1 error rate of $10^{-6}$. The population
mean $\mu=\log (1.1)=0.0953$ and power ranging from 5\%, 20\%, 50\% to
99\%. Details of the simulating parameters are given in row 3 of Table~\protect\ref{tab1}.
Each circle represents an estimate, the horizontal bar is the
averaged estimate over 200 simulated data sets, and the long horizontal
line represents the true value of $\mu$. The Bias, sample Standard
Deviation (SD) and Root Mean Squared Error (RMSE) are also provided for
each estimator.}
\label{fig3}
\end{figure}

Figure~\ref{fig3} shows the performance of the estimators when the type 1 error
rate is $10^{-6}$ and the parameter values are from row 3 of Table~\ref{tab1}.
We found that all the bias correction estimators are showing a slight
overcorrection. (Note that the scale in the $y$-axis differs between
Figures~\ref{fig2} and~\ref{fig3}.) In this setting, the results of \textbf{B.BMA} and {\bf
Ghosh} are
very similar with \textbf{B.BMA} having a smaller variance. The difference
between Figures~\ref{fig2} and~\ref{fig3} is due to the fact that the significance
threshold used is drastically different, $\alpha=0.05$ for
Figure~\ref{fig2}
and $\alpha=10^{-6}$ for Figure~\ref{fig3}, while the power of the association
study of the same SNP
is kept comparable by increasing the required sample size, $n$. As a
result, the noncentrality parameter values, $\mu/ (\sigma/\sqrt
{n})$, are not directly comparable between the two cases.

\subsection{Simulation set 2---genetic model}
 Following the setup of the simulations conducted
by \citet{ghosh2008ys}, we generated data for 500 cases and 500
controls from an additive genetic model with disease prevalence of
1\%, minor allele frequency of 0.25, and the log OR, $\beta$, ranging
from $\log(1.1)$ to $\log(2)$. The threshold value is $c=5.0$, leading to the
significance level $\alpha= 2.87\times10^{-7}$.
For each log OR value, we began by generating 200 significant data\vspace*{1pt} sets
such that the association test statistic, $\hat\beta/ \widehat{\mathit{SE}}
(\hat\beta)$, is greater than $c$, where $\hat\beta$ is the log OR
estimate obtained from the logistic regression model, and $\widehat
{\mathit{SE}} (\hat\beta)$ is the estimate of the standard error of $\hat
\beta$. Using the summary statistics,
$\hat\beta$ and $\widehat{\mathit{SE}} (\hat\beta)$,
the auxiliary information such as the sample size (we used $n=1000$)
and the threshold value of the test, we applied the Bayesian methods by
letting $\hat\mu=\hat\beta$, and $S = \hat\sigma= \widehat{\mathit{SE}}
(\hat\beta) \times\sqrt{n}$.

\begin{figure}

\includegraphics{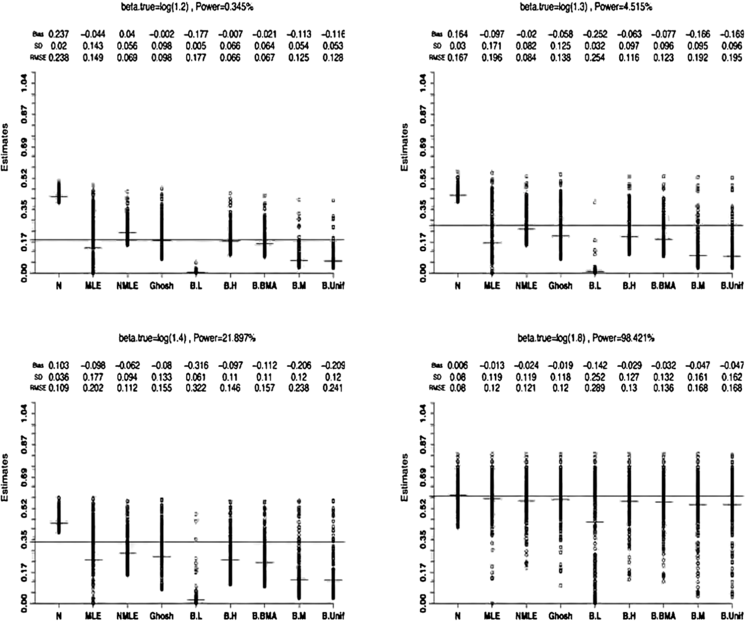}

\caption{Performance of the nine estimators under an
additive genetic model with a type 1 error rate of $\alpha
=2.87 \times 10^{-7} (c=5)$. The sample size is 1000 (500 cases and
500 controls), the minor allele frequency of the causal SNP is 0.25.
The effect of the SNP on the $\log$ OR scale ranging from $\mu=\beta=\log
(1.2)$, $\log (1.3)$, $\log (1.4)$ to $\log (1.8)$ corresponding to power
$ <$1\%, $\approx$5\%, $\approx$20\% and $>$95\% to detect the
association. Each circle represents an estimate, the horizontal bar is
the average estimate over 200 simulated data sets, and the long
horizontal line represents the true value of $\mu$. The Bias, sample
Standard Deviation (SD) and Root Mean Squared Error (RMSE) are also
provided for each estimator.}
\label{fig4}
\end{figure}

\begin{figure}

\includegraphics{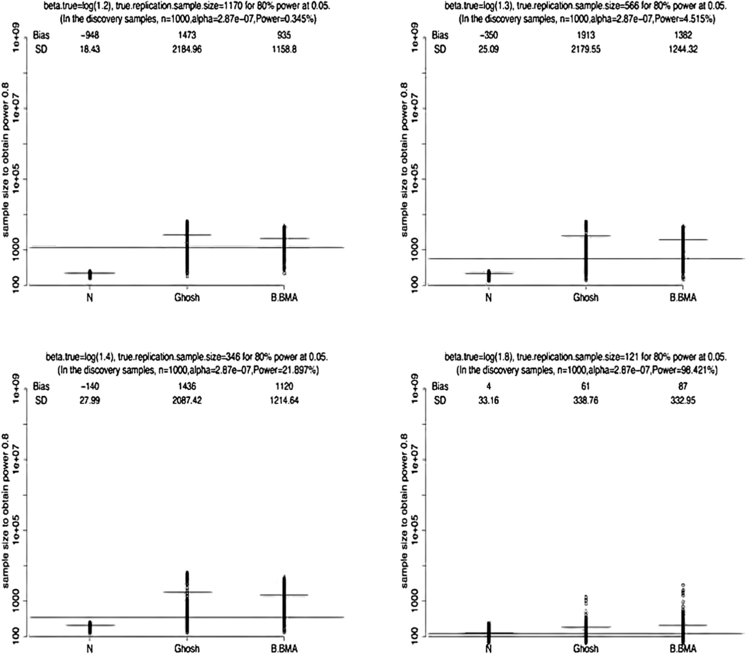}

\caption{Performance of sample size estimation for
replication studies under an additive genetic model. The initial
discovery samples are the same as those in Figure \protect\ref{fig4}.
The replication sample size is calculated assuming
a type 1 error rate of 0.05 and power of 80\%, and it is calculated
based on the estimate of the log OR by \textbf{N}, the na\"ive estimation
method, \textbf{Ghosh}, the likelihood method, or \textbf{B.BMA}, the
Bayesian method applied to the simulated significant discovery samples.
Each circle represents an estimate, the horizontal bar is the average
estimate over 200 simulated data sets, and the long horizontal line
represents the true expected required sample size.}
\label{fig5}
\end{figure}

Figure~\ref{fig4} illustrates the results for log OR values equal to
\{log(1.2), log(1.3), log(1.4), log(1.8)\}, corresponding to the power of
detecting the associated SNP in the range \{0.345\%, 4.515\%, 21.897\%,
99.5\%\}.
(Results for other log OR values are qualitatively similar.) The
results obtained from the simulated genetic models confirm that the
\textbf{B.BMA}
has a smaller RMSE than \textbf{Ghosh} when the power of the association
test is low.
Although the variance reduction on the log OR scale is small, the
implication on study design is practically important.
Figure~\ref{fig5} shows the sample size estimation for a replication study with
80\% power at the 0.05 significance level using the
na\"ive log OR estimate, the \textbf{Ghosh} estimate and the \textbf{B.BMA} estimate
obtained from the original discovery samples, as reported in Figure~\ref{fig4}.
Results show that the standard error in sample size estimation based on
\textbf{Ghosh} is almost twice as big as that based on \textbf{B.BMA} when
the power of the original association study is low (e.g., 20\% or
lower). In the low power case, we also note that the sample size
predicted based on \textbf{N}, the na\"ive estimate, is never sufficient.
For example, for a SNP with log(OR) of log(1.2), the na\"ive sample
size estimate centers around 222 with a maximum predicted size of 247,
while the true expected required sample size is 1170. Although both
\textbf{Ghosh} and \textbf{B.BMA} overestimate the necessary sample size for
replication due to the overcorrection of effect size, we believe that a
conservative sample size estimate is practically useful because it
guards against sampling variation.

We also examined different effect levels when the type I error level
is equal to 0.05 or 0.001, and we drew similar conclusions based on the
results reported in Supplement. The additional simulation studies
also include a null case where the apparent discovery is a false
positive. In that case, \textbf{B.BMA} outperforms \textbf{Ghosh}, but {\bf
B.L} performs the best, as expected.

\section{Application study}\label{sec6}

We applied the proposed Bayesian estimation methods to four data sets of
which one is a candidate gene study and the other three are genome-wide
association studies (GWAS) of either binary or quantitative outcomes.
Specifically, the four studies are as follows:
\begin{longlist}[(III)]
\item[(I)] the candidate gene association study of Lymphoma by
\citet{wangetal},
\item[(II)] the GWAS of type 1 diabetes (T1D) by \citet{wtccc},
\item[(III)] the GWAS of psoriasis by \citet{Nair2009ys},
\item[(IV)] the GWAS of complications of T1D by \citet{paterson}.
\end{longlist}
The Lymphoma and WTCCC T1D data sets were chosen because they were
previously analyzed by
\citet{ghosh2008ys} via the likelihood-based approach, and the other
two studies were chosen because the genetic effect estimates
from independent replication samples were reported by the study
authors. In
addition, the T1D complication data set allows us to demonstrate that the
proposed methods can be easily and robustly applied to association
studies of
quantitative outcomes.

In each case, the results are summarized in a table containing the
original reported genetic effect
(i.e., the na\"ive estimate, \textbf{N}), the five different Bayesian
estimators, \textbf{B.L}, \textbf{B.H}, \textbf{B.BMA}, {\bf
B.Unif} and \textbf{B.M},
and three likelihood methods, \textbf{MLE}, \textbf{NMLE} and \textbf{Ghosh}, as
described in Section~\ref{sec5}. The estimates produced by each method are
compared with the estimates obtained from the independent replication
samples reported in the literature.
We note that the anticipated power for each study differs due to the
apparent differences in study design [e.g., higher power for the
candidate gene study of \citet{wangetal} compared to the GWAS],
the sample size [e.g., higher power for the GWAS of T1D by \citet
{wtccc} with $n \approx5000$ compared to the GWAS of T1D complication
by \citet{paterson} with $n=667$], and the prior knowledge of a
SNP (e.g., higher power for rs12191877 from chromosome 6 in the
well-known MHC region that is strongly associated with Psoriasis
compared to other novel SNPs). However, we report estimates from all
five Bayesian estimators for a more complete comparison.
The estimate from the replication samples serves as the benchmark,
but the value itself should not be viewed as the true
parameter value because of the sampling variation and the potential
subpopulation and ascertainment differences between the original
discovery and
the follow-up replication studies.

We also report the corresponding confidence interval (CI) or the
highest posterior density region/interval (HpdI), but it should be
noted that the
statistical interpretations of CI and HpdI are different and,
therefore, these regions are not directly comparable.
Although the HpdI with posterior mass $1-\eta$ may be estimated using
samples from the posterior under model $M_1$ for \textbf{B.L} or $M_2$ for
\textbf{B.H}, there is no direct way to construct a HPD region for {\bf
B.BMA}, the model averaging estimator for the two models. However, a
credible interval (CrdI) can be constructed using the normal
approximation based on the model averaging estimator and its variance
estimate [see equation (7) in \citet{villa}]. For the
likelihood-based methods, we construct the CI following the method
proposed by \citet{ghosh2008ys} that was shown to outperform the
standard CI procedure. Specifically, the Ghosh $1-\eta$ CI is the
interval between the $\eta/2$ and $1-\eta/2$ quantiles of the
conditional density $p(T_n|T_n>c)$. \citet{ghosh2008ys} noted
that, although they proposed three competing point estimates, {\bf
MLE}, \textbf{NMLE} and \textbf{Ghosh}, their procedure provided only a
single CI.

\subsection{Application I---A candidate-gene study of lymphoma}
Wang  et~al. (\citeyear{wangetal}) performed a candidate gene study of Lymphoma
using a total of 48 SNPs genotyped on 318 cases and 766 controls, and
they reported two significant SNPs using a $p$-value threshold of $\alpha
= 0.002$. The na\"ive log OR estimate is
log(1.54) for rs1800629 and log(1.40) for rs909253, however, the
follow-up estimates obtained from a larger independent study
are reduced considerably to log(1.29) for rs1800629 and log(1.16) for
rs909253 [\citet{rothman}; \citet{ghosh2008ys}]. For each of the two
SNPs, we applied the likelihood estimation methods as well as the
Bayesian methods, using the na\"ive log OR estimates, $\hat\mu= \hat
\beta$, and $S=\hat\sigma= \widehat{\mathit{SE}} (\hat\beta) \times\sqrt
{n}$ inferred from the observed association $p$-value [$p$-value${}= 1- \Phi
(|\hat\beta/ \widehat{\mathit{SE}} (\hat\beta)|)$],
$n=318+766=1084$ and $c=2.878$ corresponding to $\alpha=0.002$
(Table~\ref{tab2}).

\begin{table}
\caption{Application I---the candidate gene study of Lymphoma by \protect\citet{wangetal}}\label{tab2}
\begin{tabular*}{\tablewidth}{@{\extracolsep{4in minus 4in}}lll@{}}
\hline \textbf{SNPs of interest} & \multicolumn{1}{c}{$\bolds{\mathrm{rs}1800629}$} &\multicolumn{1}{c}{$\bolds{\mathrm{rs}909253}$}\\
 \hline
 \textit{Discovery samples} & &\\
 \quad Association $p$-value&$5.7\times10^{-4}$&$ 7.4\times10^{-4}$\\
  \quad Reported effect&0.432&0.337\\[3pt]
   \textit{Likelihood estimates} & & \\
\quad MLE   (CI)&0.116   (0.000,  0.645)&0.010   (0.000,  0.498)\\
 \quad NMLE   (CI)&0.247   (0.000,  0.645)&0.184   (0.000,  0.498)\\
  \quad Ghosh   (CI)&0.182   (0.000,  0.645)&0.097   (0.000,  0.498)\\[3pt]
 \textit{Bayesian estimates} &  &\\
  \quad B.L   (HpdI)&0.005   (0.000,  0.013)&0.004   (0.000,  0.005) \\
  \quad B.H   (HpdI)&0.196   (0.000,  0.508)&0.142   (0.000,  0.382)\\
  \quad B.BMA   (CrdI)&0.150   (0.000,  0.428) &0.115   (0.000,  0.324)\\
  \quad B.Unif   (HpdI)&0.068   (0.000,  0.377)&0.045   (0.000,  0.277) \\
  \quad B.M   (HpdI)&0.074   (0.000,  0.397)&0.049   (0.000,  0.281)\\[3pt]
 \textit{Follow-up samples}  & &\\
  \quad Follow-up estimate &0.255&0.148\\
\hline
\end{tabular*}
\tabnotetext[]{}{\textit{Notes}: The
Reported Effect is na\"ive log OR estimate obtained from the original discovery samples   (318 cases and 766 controls) of \citet{wangetal},
in which the association tests of these two SNPs were significant at the $\alpha=0.002$ level. The follow-up estimate was obtained
from a larger pooled analysis by \citet{rothman}. The other eight estimates were based on either the likelihood approach,
{\bf MLE},   {\bf NMLE} and {\bf Ghosh},   or the proposed Bayesian approach,   {\bf B.L},   {\bf B.H},   {\bf B.BMA},
{\bf B.Unif} and {\bf B.M} as summarized in Section~\ref{sec5}.  CI is the 95\% confidence interval for the likelihood estimates,
 HpdI is the highest posterior
density interval with posterior mass 95\% and CrdI is the credible interval for the Bayesian estimates.}
\end{table}

Results in Table~\ref{tab2} are consistent with simulation results of power 50\%
in Figure~\ref{fig2}. Because of the anticipated high power of a candidate gene study,
both \textbf{B.BMA} and \textbf{Ghosh} overcorrect slightly with similar performance.
We observe that the CrdI of \textbf{B.BMA} is smaller than the CI of {\bf
Ghosh}, although we noted before that the interpretation of the two
intervals is different. Results suggest that \textbf{B.H} performs best
among all the Bayesian methods, which is not surprising for a study
with putative high power.

\subsection{Application II---A GWAS of Type 1 Diabetes}

The Type 1 Diabetes (T1D) GWAS from the WTCCC included approximatively
$2000$ cases and $3000$ controls and the samples were genotyped on
the Affymetrix 500K chip3 [\citet{wtccc}]. After a set of quality
control criterions (e.g., the minor allele frequency of a SNP $>5\%$,
the genotyping missing rate $<5\%$ and the $p$-value of the
Hardy--Weinberg Equilibrium test $> 5.7 \times10^{-7}$),  the authors
reported six significant loci at the $5 \times10^{-7}$ level. We
focused on the four SNPs analyzed by \citet{ghosh2008ys}
because the replication results are available from the study of
\citet{toddetal}. For each SNP of interest, we applied the
proposed estimation methods using the reported log OR estimates
obtained from the WTCCC discovery samples, $\hat\beta= \hat\mu$,
and $S=\hat\sigma= \widehat{\mathit{SE}} (\hat\beta) \times\sqrt{n}$
inferred from the observed association $p$-value, and $c=4.892$
corresponding to $\alpha=5 \times10^{-7}$ (Table~\ref{tab3}). In this
application, the actual number of cases is $1963 - 37 = 1926$ and the
number of controls is $(1480 - 24) + (1458 - 42) = 2872$, where the
37, 24 and 42 samples were deleted due to quality control issues, based
on the information provided in the supplementary Tables 1 and 4 of
\citet{wtccc}. Thus, $n=1926+2872=4798$ in this application.

\begin{sidewaystable}
\tablewidth=\textwidth
\caption{Application II---the GWAS of T1D by \protect\citet{wtccc}}\label{tab3}
\begin{tabular*}{\tablewidth}{@{\extracolsep{4in minus 4in}}lllll@{}}
\hline \textbf{SNPs of interest} & \multicolumn{1}{c}{\textbf{rs17696736}} &\multicolumn{1}{c}{\textbf{rs2292239}}
&\multicolumn{1}{c}{\textbf{rs12708716}}&\multicolumn{1}{c@{}}{\textbf{rs2542151}}\\
\hline
 {\it Discovery samples} & & & & \\
\quad Association $p$-value&$7.27\times10^{-14}$&$1.49\times10^{-9}$&\phantom{$-$}$1.28\times10^{-8}$&$ 8.4\times10^{-8}$\\
 \quad Reported effect   (CI)&0.315  (0.239,  0.399)&0.262  (0.182,  0.351)&$-$0.261  ($-$0.357,  $-$0.174)&0.285  (0.182, 0.399)\\[3pt]
 \textit{Likelihood estimates} & & & & \\
 \quad MLE  (CI)&0.314  (0.224,  0.397)&0.241  (0.095,  0.346)&$-$0.212  ($-$0.348,  0.000)&0.140  (0.000,  0.375)\\
 \quad NMLE  (CI)&0.310  (0.224,  0.397)&0.217  (0.095,  0.346)&$-$0.182  ($-$0.348,  0.000)&0.154  (0.000,  0.375)\\
 \quad Ghosh  (CI)&0.312  (0.224,  0.397)&0.229  (0.095,  0.346)&$-$0.197  ($-$0.348,  0.000)&0.147  (0.000, 0.375)\\[3pt]
 \textit{Bayesian estimates} & & & & \\
 \quad B.L (HpdI)&0.311 (0.221, 0.399)&0.019 (0.000,  0.210) &$-$0.006 ($-$0.008, 0.000) &0.004 (0.000,  0.010) \\
 \quad B.H (HpdI)&0.309 (0.221, 0.403)&0.212 (0.063, 0.345)&$-$0.170 ($-$0.306, 0.000)&0.126 (0.000,  0.294)\\
 \quad B.BMA (CrdI)&0.309 (0.234, 0.385) &0.207 (0.079, 0.336)&$-$0.161 ($-$0.318, $-$0.004)&0.117 (0.000,  0.280)\\
 \quad B.Unif (HpdI)&0.311 (0.220, 0.398)&0.172 (0.000,  0.312)&$-$0.087 ($-$0.283, 0.000)&0.045 (0.000,  0.240) \\
 \quad B.M (HpdI)&0.309 (0.211, 0.391)&0.173 (0.000,  0.310)&$-$0.092 ($-$0.286, 0.000)&0.046 (0.000, 0.249)\\[3pt]
 {\it  Follow-up samples} & & & & \\
 \quad Follow-up estimate  (CI)&0.148  (0.086,  0.207)&0.247  (0.182,  0.308)&$-$0.186  ($-$0.248,  $-$0.116)&0.254  (0.174,  0.337)\\
\hline
\end{tabular*}
\tabnotetext[]{}{\textit{Notes}: The
reported effect is na\"ive log OR estimate obtained from the original discovery samples    (1926 cases and 2872 controls) of \citet{wtccc},
  in which the association tests of these SNPs were significant at the $\alpha=5 \times 10^{-7}$ level. The Follow-up Estimate was obtained
  from the replication study by \citet{toddetal}.
The other eight estimates were based on either the likelihood approach,   {\bf MLE},   {\bf NMLE} and {\bf Ghosh},   or the proposed Bayesian
approach,   {\bf B.L},   {\bf B.H},   {\bf B.BMA},   {\bf B.Unif} and {\bf B.M} as summarized in Section~\protect\ref{sec5}.  CI is the 95\% confidence interval
 for the likelihood estimates,   HpdI is the highest posterior
density interval with posterior mass 95\% and CrdI is the credible interval for the Bayesian estimates.}
\end{sidewaystable}

Results in Table~\ref{tab3} show that if the original association result is
extreme in that the $p$-value is considerably smaller than the threshold
considered (i.e., rs17696736), then the prior influences the result
only minimally. Similarly, the likelihood-based estimates are only
slightly reduced from the published estimated log ORs. However, the
follow-up estimate is considerably lower than the bias reduced
estimates. As noted by \citet{ghosh2008ys}, this suggests
possible heterogeneity between the discovery and replication samples. A
subtle but important explanation for the results in the last three
columns of Table 3 where the replicated values are larger in absolute
value than the estimates produced by each method is that the follow-up
estimates here are also subject to the winner's curse, albeit less
severe, because only estimates of successfully replicated SNPs were reported.

\begin{sidewaystable}
\tablewidth=\textwidth
\caption{Application III---the GWAS of Psoriasis by \protect\citet{Nair2009ys}}\label{tab4}
\begin{tabular*}{\tablewidth}{@{\extracolsep{4in minus 4in}}llllll@{}}
\hline \textbf{SNPs of interest} & \multicolumn{1}{c}{\textbf{rs12191877}} &\multicolumn{1}{c}{\textbf{rs2082412}}
&\multicolumn{1}{c}{\textbf{rs17728338}}&\multicolumn{1}{c}{\textbf{rs20541}}&\multicolumn{1}{c@{}}{\textbf{rs610604}}\\
\hline
 {\it Discovery samples} & & & &  & \\
\quad $p$-value&$4\times10^{-53}$&$5\times10^{-10}$&$2\times10^{-7}$&$ 6\times10^{-6}$&$1\times10^{-5}$\\
 \quad Reported effect &1.026&0.445&0.542&0.315&0.247\\[3pt]
\textit{Likelihood estimate} & & & &  & \\
 \quad MLE  (CI)&1.026  (0.895,  1.157)&0.443  (0.287,  0.585)&0.514  (0.214,  0.746)&0.234  (0.000,  0.445)&0.162  (0.000,  0.349)\\
 \quad NMLE  (CI)&1.026  (0.895,  1.157)&0.435  (0.287,  0.585)&0.476  (0.214,  0.746)&0.210  (0.000,  0.445)&0.154  (0.000,  0.349)
 \\
 \quad Ghosh  (CI)&1.026  (0.895,  1.157)&0.439  (0.287,  0.585)&0.495  (0.214,  0.746)&0.222  (0.000,  0.445)&0.158  (0.000, 0.349)\\[3pt]
 \textit{Bayesian estimate} & & & &  & \\
 \quad B.L (hpdI)&1.026 (0.887,  1.153)&0.400 (0.000,  0.556) &0.049 (0.000,  0.494)&0.007 (0.000,  0.010)&0.005 (0.000,  0.009) \\
 \quad B.H (hpdI)&1.024 (0.891, 1.150)&0.436 (0.276, 0.587)&0.468 (0.170, 0.754)&0.197 (0.000,  0.377)&0.136 (0.000,  0.288)\\
 \quad B.BMA (CrdI)&1.024 (0.915, 1.132)&0.436 (0.304, 0.568)&0.444 (0.151, 0.738)&0.172 (0.000,  0.379)&0.122 (0.000,  0.279)\\
 \quad B.Unif (hpdI)&1.026 (0.898, 1.163) &0.437 (0.283, 0.592)&0.405 (0.000,  0.681)&0.094 (0.000,  0.339)&0.062 (0.000,  0.252) \\
 \quad B.M (hpdI)&1.026 (0.887, 1.146) &0.436 (0.268, 0.580)&0.402 (0.000,  0.687)&0.096 (0.000,  0.341)&0.063 (0.000,  0.253)\\[3pt]
 {\it Follow-up samples} & & & &  &\\
 \quad Follow-up estimate &0.971&0.365&0.464&0.239&0.174\\
 \hline
 \end{tabular*}
 \end{sidewaystable}
 \setcounter{table}{3}
\begin{sidewaystable}
\tablewidth=\textwidth
\caption{(Continued)}
\begin{tabular*}{\tablewidth}{@{\extracolsep{4in minus 4in}}llllll@{}}
\hline
 \textbf{SNPs of interest} & \multicolumn{1}{c}{\textbf{rs2066808}} &\multicolumn{1}{c}{\textbf{rs2201841}}&\multicolumn{1}{c}{\textbf{rs1076160}}&\multicolumn{1}{c@{}}{\textbf{rs12983316}}\\
  \hline
 {\it Discovery samples} & & & &  \\
 \quad Association $p$-value&$2\times10^{-5}$&$3\times10^{-7}$&$2\times10^{-5}$&$2\times10^{-5}$\\
 \quad Reported effect &0.519&0.300&0.231&0.308\\[3pt]
 \textit{Likelihood estimates} & & & &  \\
 \quad MLE  (CI)&0.231  (0.000,  0.728)&0.281  (0.107,  0.414)&0.103  (0.000,  0.324)&0.137  (0.000,  0.432)\\
\quad NMLE  (CI)&0.293  (0.000,  0.728)&0.258  (0.107,  0.414)&0.129  (0.000,  0.324)&0.173  (0.000,  0.432) \\
 \quad Gho0sh  (CI)&0.262  (0.000,  0.728)&0.270  (0.107,  0.414)&0.116  (0.000,  0.324)&0.155  (0.000, 0.432)\\[3pt]
 \textit{Bayesian estimates} & & & &  \\
 \quad B.L  (HpdI)&0.008 (0.000,  0.011)&0.021 (0.000,  0.228)&0.003 (0.000,  0.005)&0.004 (0.000,  0.010) \\
 \quad B.H  (HpdI)&0.247 (0.000,  0.571)&0.253 (0.076, 0.422)&0.110 (0.000,  0.257)&0.147 (0.000,  0.340)\\
 \quad B.BMA  (CrdI)&0.221 (0.000,  0.54) &0.240 (0.074, 0.407)&0.097 (0.000,  0.239)&0.127 (0.000,  0.316)\\
 \quad B.Unif  (HpdI)&0.097 (0.000,  0.472)&0.207 (0.000,  0.381)&0.042 (0.000,  0.209)&0.056 (0.000,  0.275) \\
 \quad B.M (HpdI)&0.099 (0.000,  0.482)&0.210 (0.000,  0.376)&0.044 (0.000,  0.213)&0.057 (0.000,  0.273)\\[3pt]
 {\it Follow-up samples} & & & &  \\
 \quad Follow-up estimate &0.293&0.122&0.086&0.086\\
\hline
\end{tabular*}
\tabnotetext[]{}{\textit{Notes}: The reported effect is na\"ive log OR estimate obtained from the original discovery samples
(1359 cases and 1400 controls) of \citet{Nair2009ys},   in which  these SNPs were among the top 2000 SNPs based
on the p-values of the association tests,   corresponding to $\alpha=10^{-4}$ level. The Follow-up estimate was obtained
from the replication study by \citet{Nair2009ys}.
The other eight estimates were based on either the likelihood approach,   {\bf MLE},   {\bf NMLE} and {\bf Ghosh},
or the proposed Bayesian approach,   {\bf B.L},   {\bf B.H},   {\bf B.BMA},   {\bf B.Unif} and {\bf B.M} as summarized
in Section~\ref{sec5}.  CI is the 95\% confidence interval for the likelihood estimates,   HpdI is the highest posterior
density interval with posterior mass 95\% and CrdI is the credible interval for the Bayesian estimates.}
\end{sidewaystable}

\subsection{Application III---A GWAS of Psoriasis}
 Nair, Duffin and Helms (\citeyear{Nair2009ys})  conducted a two-stage association of
Psoriasis, a chronic skin disease characterized by circumscribed red
patches covered with white scales.
The first stage is a GWAS with 438,670 SNPs genotyped on 1359 cases
and 1400 controls, and the second stage is a replication study
following up on 21 promising SNPs using a set of independent 5048
cases and 5051 controls.
{\it``Owing to the winner's curse, odds ratios estimated in the
discovery sample were
larger than those estimated in the follow-up samples''} [Table 2 of
\citet{Nair2009ys}]. The SNP selection criterion was mainly based
on the ranking of the GWAS $p$-value, roughly corresponding to a $p$-value
threshold of $\alpha=10^{-4}$. For each SNP of interest, we applied
the estimation methods using the reported log OR estimates obtained
from the discovery samples, $\hat\beta= \hat\mu$, and $S=\hat
\sigma= \widehat{\mathit{SE}} (\hat\beta) \times\sqrt{n}$ inferred from
the observed association $p$-value,
$n=1359+1400=2759$ and $c=3.719$ corresponding to $\alpha=10^{-4}$
(Table~\ref{tab4}).

When the results are as extreme as rs12191877 with $p=4\times10^{-53}$
or as rs2082412 with $p=5\times10^{-10}$, indicating high power at the
chosen threshold level, all the bias correction estimators results in
little change from
the published estimate, including \textbf{B.L} despite its inherent prior
skepticism of a finding. For
the other less significant SNPs in the table, both \textbf{B.BMA} and {\bf
Ghosh} achieve substantial bias reduction.
In general, \textbf{B.BMA} has a noticeably smaller variance for lower
power cases,
which in turn can produce more reliable sample size estimates for
replication studies.

\subsection{Application IV---A GWAS of quantitative measures of T1D
complications}
In the fourth setting of the GWA study of
longitudinal repeated quantitative measures of phenotype HbA1c in
the Diabetes Control and Complications Trial (DCCT) samples, a
significant locus (at $\alpha=5 \times10^{-8}$) was identified in
the conventional treatment group with 667 samples near SORCS1
(rs1358030 with $p$-value${} = 4.66\times10^{-9}$).
The association statistic was obtained via regression analysis
of the average log (HbA1c) value vs. SNP with an additive genotype
coding. The GWAS was performed on 841,342 SNPs, genotyped by the
Illumina 1M BeadArray assay, that passed a set of quality control
criteria [details in \citet{paterson}].

\begin{table}[b]
\tablewidth=250pt
\caption{Application IV---the GWAS of HbA1c in Type 1 Diabetes patients,
by \protect\citet{paterson}}\label{tab5}
\begin{tabular*}{250pt}{@{\extracolsep{4in minus 4in}}ll@{}}
\hline \textbf{SNP of interest} & \textbf{rs1358030}\\ \hline
 {\it Discovery samples} & \\
 \quad Association $p$-value&$4.66\times10^{-9}$\\
 \quad Reported effect & 0.045\\[3pt]
 \textit{Likelihood estimates} & \\
 \quad MLE   (CI)&0.029  (0.000,  0.056)\\
 \quad NMLE   (CI)&0.024   (0.000,  0.056)\\
 \quad Ghosh   (CI)&0.027   (0.000,  0.056)\\[3pt]
 \textit{Bayesian estimates} &\\
 \quad B.L  (HpdI)&0.001   (0.000,  0.002) \\
 \quad B.H   (HpdI)&0.021   (0.000,  0.048)\\
 \quad B.BMA   (CrdI)&0.020   (0.000,  0.047)\\
 \quad B.Unif   (HpdI)&0.007   (0.000,  0.040) \\
 \quad B.M   (HpdI)&0.008   (0.000,  0.040)\\[3pt]
  {\it Follow-up samples} & \\
 \quad Follow-up estimate &0.005\\
\hline
\end{tabular*}
\tabnotetext[]{}{\textit{Notes}: The reported effect is the na\"ive estimate of the regression coefficient obtained from the 667 discovery samples,
in which the association test of the SNP was significant at
the  $\alpha=5 \times 10^{-8}$ level. The Follow-up estimate was obtained from 637 independent samples.
The other eight estimates were based on either the likelihood approach,   {\bf MLE},   {\bf NMLE} and {\bf Ghosh},
 or the proposed Bayesian approach,   {\bf B.L},   {\bf B.H},   {\bf B.BMA},   {\bf B.Unif} and {\bf B.M} as summarized in Section~\ref{sec5}.
  CI is the 95\% confidence interval for the likelihood estimates,   HpdI is the highest posterior
density interval with posterior mass 95\% and CrdI is the credible interval for the Bayesian estimates.}
\end{table}

The na\"ive estimate of the regression coefficient for
rs1358030 is 0.045. However, the estimate obtained from the
intensive treatment
group with 637 samples is 0.005 (Table~\ref{tab5}). Note that for the intensive
treatment group, only the measures at the eligibility time-point
(i.e.,\ before the starting of the two different treatments) were used
for the regression analysis so that the two groups are
comparable and the intensive treatment group could be used as a
replication data set.

Unlike the case control studies with binary response (diseased or not)
considered previously, of interest here is a quantitative outcome,
HbA1c, that measures the amount of glycated hemoglobin in blood.
Therefore, the $\mu$ no longer represents the log OR but the
corresponding coefficient in the linear regression model. Although we
could consider choosing a more suitable prior, we adopted the same
$\operatorname{Uniform}(0, 2)$ density
for $f (\mu)$ as for the case-control data to test the robustness of
the Bayesian methods. (Results from other prior choices are discussed
in Section~\ref{sec7}.) To apply the Bayesian methods, we let $\hat\mu=0.045$,
$n=667$, $c=5.328$ (corresponding to the threshold used, the
significance level is $\alpha=5 \times10^{-8}$), and the observed
association $p$-value $4.66\times10^{-9}$ (corresponding to a test
statistic of 5.743) allows us to infer the standard error $S=\mu*\sqrt
{n}/5.743=0.202$ (Table~\ref{sec5}). As expected for the low power case, both
\textbf{B.BMA} and \textbf{Ghosh} reduce the estimation bias but not
sufficiently enough, and \textbf{B.L} performs better. However, in this
case the estimates from \textbf{B.Unif} or \textbf{B.M} are closest to the
one obtained from the follow-up study.

\section{Conclusions and future work}\label{sec7}

We propose hierarchical Bayes methods to reduce selection bias in
genetic association studies. The basis of the approach is a
spike-and-slab prior which essentially allows for the possibility that
the signal detected may be a false positive. The prior
permits the researchers to quantify their belief in the strength of the
signal. Depending on the prior, inference based on the posterior
distribution may be different from model to model and, therefore, the
researcher faces a (sometimes difficult) choice. To alleviate this
dilemma, we consider a Bayesian model averaging strategy, \textbf{B.BMA},
in which we use the data to weigh in on the more appropriate model.

Simulation and application studies demonstrated that the \textbf{B.BMA}
estimator performs well across different settings, and we recommend
\textbf{B.BMA} when there is little information on the putative power of
the initial discovery study.
However, we also emphasize that model averaging is not necessarily the
best approach for a given study. Factors such as study design and
sample size should be taken into account in the decision of using a
more conservative model like \textbf{B.L} or an anti-conservative one like
\textbf{B.H}. In general, \textbf{B.H} is suitable for candidate gene studies
with putative high power as demonstrated in application I, and {\bf
B.L} is preferred for GWAS with putative low power as shown in
application IV. Knowledge about the SNP of interest is also a factor.
For example, little bias is expected for a SNP in a well-known
associated region or with $p$-value significantly smaller than the chosen
threshold as demonstrated by the first SNP (rs12191877) in Table 4 of
application III, while substantial bias is expected for a SNP with
$p$-value just below the threshold as shown by the last SNP (rs12983316)
in the table.

We have carried out additional simulation studies to investigate the
robustness of the Bayesian estimators. Results provided in
\hyperref[sup]{Supplement}
show that the proposed methods are robust to the choice of prior for
$\xi$, the hyperparameter that reflects our prior belief in false
positive, to the number of iterations discarded from the MCMC sample,
and to the value of $A$, the prior upper bound of log odds ratio. In addition,
we developed our methods using a conceptual normal model but
demonstrated via simulations and applications that this normal model is
well connected with widely used real genetic models and is robust to
the choice of priors.
For example, in application IV when the phenotype is not a case-control
status but a quantitative outcome, we kept the same $A=2$ knowing that the
the upper bound for $\mu$, the genetic effect size, in this case can
be reasonably assumed to be 0.2. To be more precise, note that $\mu$
is a regression coefficient in this setup and is related
to the percentage of phenotype variation explained by the SNP via the expression
\[
r^2=\mu^2 \frac{S_X^2}{S_Y^2},
\]
where $S_X^2 \approx0.467$ is
the sample variance of the SNP and $S_Y^2 \approx0.018$ is the sample
variance of the phenotype.
Since $r^2 \leq100\%$, thus, $\mu\leq0.2$. When $A=0.2$ was assumed,
the estimates were largely unchanged
compared to results in Table 5: 0.00062 (0, 0.001) for \textbf{B.L},
0.021 (0.000, 0.0474) for \textbf{B.H}, 0.0197 (0, 0.0456) for {\bf
B.BMA}, 0.0077 (0.000, 0.03996) for \textbf{B.Unif} and 0.0084 (0.000,
0.0407) for \textbf{B.M}.
If a true effect is greater than 2, our Bayesian estimations will be
bounded by 2. In practice, if the true OR is greater than $\exp(2)
\approx7.4$,
then the putative power of the original association study is very high
(unless the sample size is extremely small), resulting in little
estimation bias of the na\"ive estimate. Second, if a Bayesian estimate
was close to the upper bound, then one can choose a bigger value such
as 6.
This modification does not affect the estimation for the cases when the
effects are less than 2 (confirmed by our additional simulation
studies) but provide better effect estimates when the true effects are
indeed greater than 2. The proposed Bayesian methods, however, are not
robust to the misspecification of the threshold used. This type of
sensitivity was also observed for other existing methods including the
likelihood and resampling based methods.

The \textbf{NMLE} estimator proposed by \citet{ghosh2008ys} is the
mean of the normalized conditional
likelihood, and it can be interpreted as the posterior mean with an
improper flat prior on $\mu$ which should produce similar results
to \textbf{B.Unif}. However, unlike \textbf{NMLE}, our model allows a point
mass on effect being equal to 0 via the spike-and-slab prior, leading
to a better performance than \textbf{NMLE}. As an
average of the conditional \textbf{MLE} and the \textbf{NMLE} estimators, the
\textbf{Ghosh} estimator strikes a balance between the two
and performs better than both across different settings. Although {\bf
Ghosh} and \textbf{B.BMA} can have similar performance in some settings,
the advantage of the proposed Bayesian estimator is clear and
meaningful. For example, the standard error in sample size estimation
based on \textbf{B.BMA} is almost twice as small as that based on {\bf
Ghosh} when the power of the original association study is low as shown
in Figure~\ref{fig5}.

Both the likelihood and Bayesian methods correct for threshold effect
(i.e., the SNP of interest must pass a significance threshold) by
incorporating the threshold value in the models. In practice, another
source of bias is the ranking effect.
More precisely, suppose that a large number of SNPs are considered but
only the effects for top ranked SNPs are estimated.
Again, the effect estimate is biased but a likelihood-based correction
is cumbersome since all SNPs (with complex correlation structure among
them due to linkage disequilibrium) must be considered jointly. The
proposed Bayesian method only indirectly models the ranking effect by
allowing the SNP of interest to be false positive. So far, the method
of choice for this problem remains the bootstrap-based correction
method of \citet{Sun2005ys}.
However, the bootstrap method requires the original individual specific
data which can be limiting. In contrast,
the Bayesian and the likelihood approaches only need the summary
statistics such as the reported na\"ive estimate and the association
$p$-value, and the auxiliary information such as the sample size and the
threshold used.
In a two-stage setting when both the original discovery scan and a
replication study are available, the combined approach proposed by
\citet{Bowden2009ys} could provide better estimation results.

Although the method proposed here falls within the Bayesian paradigm, it
has a clear frequentist component since the sampling distribution is
conditional on the significance of the hypothesis test. While a
complete Bayesian analysis in which simultaneous testing and estimation is
possible for the problems considered here, it must be noted that the
current practice among genetic investigators is to perform a large
number of individual association tests prior to moving on to the
estimation stage, in part due to the computational challenges
associated with analyzing 500,000 or more SNPs.
It is for this reason and to address the bias incurred by the resulting
inference that we chose to use the current model. A full joint Bayesian
analysis is the subject of ongoing research.

\section*{Acknowledgments}
We would like to thank the Editor, an
Associate Editor and three reviewers for constructive comments and
suggestions that have substantially improved the paper. We would also
like to thank Dr. Andrew Paterson for insightful discussions of the
association study of complications in type 1 diabetes patients.

\begin{supplement}[id=suppA]\label{sup}
\sname{Supplement}
\stitle{Additional Derivations and Simulation Plots\\}
\slink[doi,text={10.1214/ 10-AOAS373SUPP}]{10.1214/10-AOAS373SUPP}
\slink[url]{http://lib.stat.cmu.edu/aoas/373}
\sdatatype{.pdf}
\sdescription{The appendix contains derivations related to posterior
computation and additional simulation results related to the robustness
of the Bayesian model considered to the choice of prior. }
\end{supplement}


\printaddresses

\end{document}